\documentclass[prl,twocolumn,showpacs,preprintnumbers,amsmath,amssymb]{revtex4}
\usepackage{graphicx,lscape}
\tighten
\begin{document}
\title{
Nonlinear dynamics of polariton scattering in semiconductor microcavity:
bistability vs stimulated scattering
}
\author{N.~A.~Gippius,$^a$  S.~G.~Tikhodeev,$^a$
V.~D.~Kulakovskii,$^b$ D.~N.~Krizhanovskii,$^b$ and
A.~I.~Tartakovskii$^b$
}
\affiliation{
(a) General Physics Institute, Russian Academy of Sciences, Vavilova 38,
Moscow 119991, Russia\\
(b) Institute of Solid State Physics, RAS, Chernogolovka, 142432 Russia
}


\begin{abstract}
We demonstrate experimentally an unusual behavior of the parametric
polariton scattering in semiconductor microcavity under a strong cw
resonant excitation. The maximum of the scattered  signal above the
threshold of stimulated parametric scattering does not shift along the
microcavity lower polariton branch with the change of pump detuning or
angle of incidence but is stuck around the normal direction. We show
theoretically that such a behavior can be modelled numerically by a
system of Maxwell and nonlinear Schr\"{o}dinger equations for cavity
polaritons and explained via the competition between the bistability of
a driven nonlinear MC polariton and the instabilities of parametric
polariton-polariton scattering.
\end{abstract}

\maketitle

The effect of parametric scattering of semiconductor microcavity
(MC) polaritons has been established
recently~\cite{Savvidis00,Houdre00,Ciuti00,Tartakovskii00,
Stevenson00,Baumberg00,Ciuti01,Whittaker01,Savvidis01,Saba01,Savasta03,Huynh03}
and attracted a significant interest. Extremely large parametric
gains~\cite{Savvidis00,Saba01} and striking transformations of the
polariton parametric luminescence with the  increase of intensity
of resonant coherent excitation~\cite{Houdre00,Tartakovskii00,
Stevenson00,Baumberg00,Savvidis01} have been demonstrated and
analyzed
theoretically~\cite{Ciuti00,Ciuti01,Whittaker01,Savvidis01,Savasta03}.
However, only the case of specific pump conditions known
as \textit{magic angle} excitation, when the parametric scattering
threshold intensity is minimum, has been investigated so far in
detail in case of stationary excitation.

In this Letter we investigate the parametric MC
polariton-polariton scattering varying the cw pump angle and
energy away from the magic angle conditions. According to the
existent theory~\cite{Ciuti01,Whittaker01,Savvidis01},
the maxima of scattered  signal
have to shift along the blueshifted and renormalized lower polariton (LP)
branch. To the
contrary, we found experimentally that  above the threshold of
stimulated parametric scattering the signal maximum stays
approximately in the normal direction, notwithstanding the change
of the excitation conditions~\cite{Kulakovskii01-2}.

We show numerically that a qualitatively similar behaviour can be
demonstrated by a system of Maxwell and nonlinear Schr\"{o}dinger
equations for MC quantum well excitons, accounting for the
exciton-cavity photon coupling into MC polariton and
polariton-polariton scattering. In the real life,  many other
scattering processes can play a role, e.g., polariton-phonon
scattering~\cite{Piermarocchi96}, polariton-free carrier
scattering~\cite{Malpuech02}. But it turns out that the
polariton-polariton scattering alone demonstrates already sharp
transformation of the scattering process, which results in the jump
of the scattered signal from the oblique to normal direction, in a
qualitative agreement with the experiment.

In the part of polariton-polariton scattering, our description
is analogous to that in~\cite{Ciuti01,Whittaker01,Savvidis01}.
As in~\cite{Whittaker01}, we neglect the saturation term,
responsible for the Pauli principle in exciton-photon interaction.
It can be shown that its influence is not strong in case
of MC with our parameters. However, one
difference is important.  We treat the amplitude of the driven
polariton mode not as a fixed external parameter, but as an
internal parameter established  in the nonlinear system under a
fixed external pump. Well known in the theory of driven nonlinear
optical cavities without a polariton resonance (see, e.g.,
in~\cite{Firth96,Kuszelewicz00}), this description allows
to explain bistability, symmetry breaking, pattern formation,
bifurcations to chaos. As we show in this Letter, the bistability
can play an important role in the nonlinear kinetics of MC
polariton scattering as well.

\begin{figure}[t]
\vspace{-0.5cm}
\includegraphics[width=\linewidth]{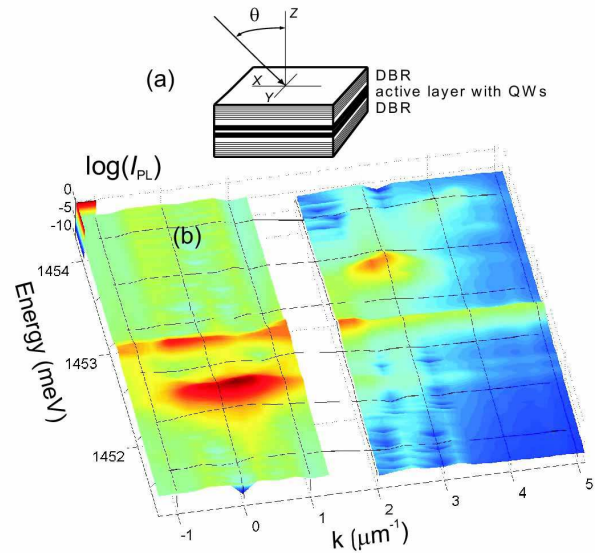}
\vspace{-0.8cm}
\caption{
\label{fig:1}The experimental
 geometry (\textit{a});
 measured energy-momentum spectra of the LP photoluminescence
 under a resonant excitation at $\hbar \Omega_0 = 1452.7$~meV
 and $\vartheta = 12.5^\circ$ with intensity above the scattering
 threshold (\textit{b}).
}
\vspace{-0.5cm}
\end{figure}

\begin{figure}[t]
\vspace{-1.0cm}
\includegraphics[width=\linewidth]{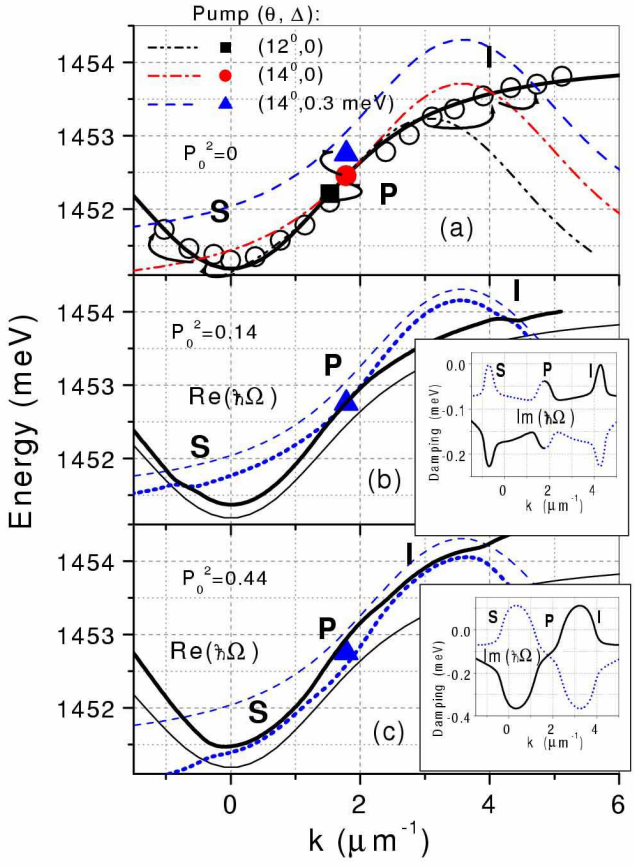}
\vspace{-1cm}
\caption{
\label{fig:2}
Lower polariton branches: measured (circles) and calculated for
$|\mathcal{P}_0|^2=0$~(\textit{a}), 0.14~(\textit{b}),
and 0.44~(\textit{c}).
Inserts in panels \textit{b, c} show the corresponding dampings.
 The lines are: signal (solid) and idler (dashed).
 Dash-dot-dot, dash-dot and dash lines in panel \textit{a}
 are idlers calculated for $|\mathcal{P}_0|^2=0$ and
 different pump angles  and  detunings $(\vartheta,  \Delta) =
 (12^\circ, 0)$ - magic excitation, (14$^\circ$,0) and (14$^\circ$, 0.3~meV).
[$\Delta = \hbar\Omega_{0}-E_{\mathrm{LP}}(\mathbf{k}_0). $]
 Square, circle,
 and triangle are the corresponding pump (P) locations.
 Thin lines in panels \textit{b, c} are  signal and idler at
 $|\mathcal{P}_0|^2=0$ from panel \textit{a}.
}
\vspace{-0.5cm}
\end{figure}

Experimentally, we have investigated photoluminescence (PL) from
GaAs/AlAs MC containing six 10-nm-thick InGaAs quantum wells
(QW's) in 3/2$\lambda$ GaAs cavity with cavity mode (cavity mode
-- QW exciton detuning -0.5~meV), see the experimental geometry in
Fig.~\ref{fig:1}\textit{a}. A sample was mounted in a cryostat
with a wide angular access (70$^o$) at a controlled temperature of
T=1.8-20 K. The experiment was done under excitation with a
tunable cw Ti-Sapphire laser under a changeable  angle of
incidence $\vartheta_0$. The angle resolved measurements of the MC
emission and the use of cw excitation have allowed to fix both the
energy and wavevector of the emission with a high precision. The
details on the sample and experimental set-up can be found
in~\cite{Tartakovskii00}.

Figure~\ref{fig:1}\textit{b} shows the distribution of the scattered
light vs. energy $\hbar \omega$ and in-plane momentum $k = \hbar \omega
\sin \vartheta / c$. The  excitation is at $\vartheta_0 = 12.5^\circ$
and $\hbar\Omega_0=1452.7 $~meV ($k_0 \approx 1.6 \mu\mathrm{m}^{-1}$),
with high excitation power of 880~W/cm$^2$, which is above the threshold
of stimulated parametric scattering. Figure shows  two pronounced
emission maxima at $k \gtrapprox 0$ and $ k \lessapprox 2k_0$. A closer
look shows that this cannot be explained within the existent theory of
the MC parametric scattering~\cite{Ciuti01,Whittaker01,Savvidis01}, even if the LP
spectra renormalization is taken into account.

The bare LP dispersion $E_{\mathrm{LP}}(\mathbf{k})$ has been
determined experimentally from the angle resolved PL measurements
under a weak above band gap excitation, see in
Fig.~\ref{fig:2}\textit{a}. The inflexion point of the LP branch is at
$\approx12^\circ$, which corresponds to momentum $\approx1.5\cdot
10^4$~cm$^{-1}$ and energy $1452.2$~meV.

At low excitation intensity
the LP polariton-polariton parametric scattering should take place
with the energy and in-plane momentum conservation,
$ \mathbf{k}+\mathbf{k}'=2\mathbf{k}_{0},\ \
E_{\mathrm{LP}}(\mathbf{k}) + E_{\mathrm{LP}}(\mathbf{k}') =2
\hbar\Omega_{0}.$
Two sharp emission peaks, \textit{signal} (S) at $\mathbf{k}_s $
and \textit{idler} (I)  at $\mathbf{k}_i$, are predicted
at the intersection of the signal and idler branches
$E_{\mathrm{LP}}(\mathbf{k})$  and
$2 \hbar\Omega_{0}- E_{\mathrm{LP}}(2\mathbf{k}_{0} - \mathbf{k})$,
see in Fig.\ref{fig:2}\textit{a}.
Under a weak pump into the inflexion point (magic angle)
the signal peak is predicted at $k_s \approx 0$.
With the shift of the pump angle and energy away from the inflection
point
S and I peaks shift along the LP branch towards $k_s < 0, \,\,k_i > 2 k_0$
as shown by  arrows in Fig.~\ref{fig:2}\textit{a}.

With increasing the pump intensity the  signal and idler branches
become renormalised via the parametric coupling~\cite{Ciuti00,Ciuti01}, see
in Fig.~\ref{fig:2}\textit{b}. Simultaneosly the dampings of signal
and idler branches at the peak positions approach zero (see insert
in panel \textit{b}). At some critical pump intensity the damping
changes sign (becomes \textit{gain}), which signals the instability
and means the threshold of the parametric stimulated scattering.
However, the prediction within this theoretical model is still $k_s
< 0, \,\,k_i > 2 k_0$, whereas on the experiment we observe the
signal and idler maxima at $k_s \gtrapprox 0, \,\,k_i \lessapprox
2k_0$.

In order to model the polariton scattering dynamics, we solve
numerically the system of coupled equations for
$\mathcal{E}_\mathrm{QW}$, the electric field on the QW,  and
$\mathcal{P}(k,t)$, the exciton polarisation integrated over the
QW thickness,
\begin{eqnarray}
 & \!\!\!\!\left[i \hbar \frac{d}{dt} \right. -& \!\! \left.
 E_\mathrm{MC} \right] \mathcal{E}_\mathrm{QW}(k,t)  =
\alpha\mathcal{E}_\mathrm{ext}(k,t) 
 + \beta\mathcal{P}(k,t),
\label{field}\\
\label{pol} &\!\!\!\!\left[i \hbar \frac{d}{dt}  \right. - & \!\! \left.
E_{X} \right] \mathcal{P}(k,t)   =
A \mathcal{E}_\mathrm{QW}(k,t)+ \xi(t)\\ \nonumber
& & + F
\sum_{q,q'}\mathcal{P}(q,t)\mathcal{P}(q',t)\mathcal{P}^\ast(q+q'-k,t).
\end{eqnarray}
 Here $\mathcal{E}_\mathrm{ext} = \mathcal{E}_\mathrm{ext}(t)
e^{-i\Omega_0t}\delta_{k,k_0}$ is the external
electric field of the incoming pump far from the MC which is
treated as coherent macro-occupied mode with fixed
$\hbar\Omega_0$ and $k_0$.
$\mathcal{E}_\mathrm{ext}(t)$  is the pump amplitude,
changing slowly with time. $E_\mathrm{MC}(k)$ and $E_{X}$ are
empty microcavity and QW exciton frequencies. $F$ and $A$ are the
 QW exciton interaction constant and susceptibility,
$\alpha(k)$ and $\beta(k)$  are
the MC response coefficients.
In these notations,
the MC polariton Rabi frequency is
$\sqrt{A\beta}$.
Both $E_\mathrm{MC}(k)$ and $E_{X}$ are
complex, with imaginary parts accounting  for finite
lifetimes of the MC mode and QW exciton.
$\xi(k,t)$ is the stochastic Langevin force.
$E_\mathrm{MC}(k)$, $\alpha(k)$, and $\beta(k)$ are calculated
via the scattering matrices of the upper and lower MC Bragg
mirrors. $E_X$ and $A$ are fitted parameters. The energy is
measured in meV, the units of polarisation and electric field
are such that $F=1$. The in-plane momentum  is two-dimensional,
but we solve numerically only a simplified one-dimensional
model.

Equation (\ref{field}) is the Maxwell equation with the exciton
polarization in the resonance approximation for the cavity.
Equation (\ref{pol}) is the inhomogeneous nonlinear
Schr\"{o}dinger equation with damping and two types of sources:
coherent external pump and stochastic Langevin noise.
Eqs.(\ref{field},\ref{pol}) allow to describe the influence of
coherent pump and quantum fluctuations simultaneously in a very
convenient for numerical modelling way.

\begin{figure}[t]
\vspace{-0.5cm}
\includegraphics[width=\linewidth]{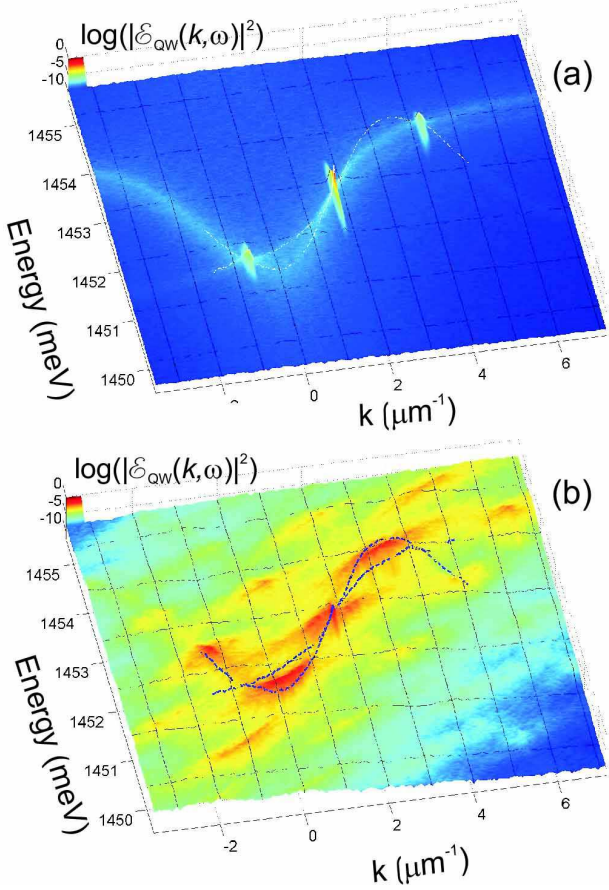}
\vspace{-0.5cm} \caption{
\label{fig:3} The calculated time-integrated energy-momentum
spectra of the scattered light from MC when the pump pulse
amplitude is slightly below (\textit{a}) and above (\textit{b})
the threshold. Dotted lines are the signal and idler branches
from Fig.~\ref{fig:2}\textit{b} calculated at the parametric
scattering threshold $|\mathcal{P}_0|^2= 0.14$.
 }
 \vspace{-0.5cm}
\end{figure}

\begin{figure}[ht]
\includegraphics[width=\linewidth]{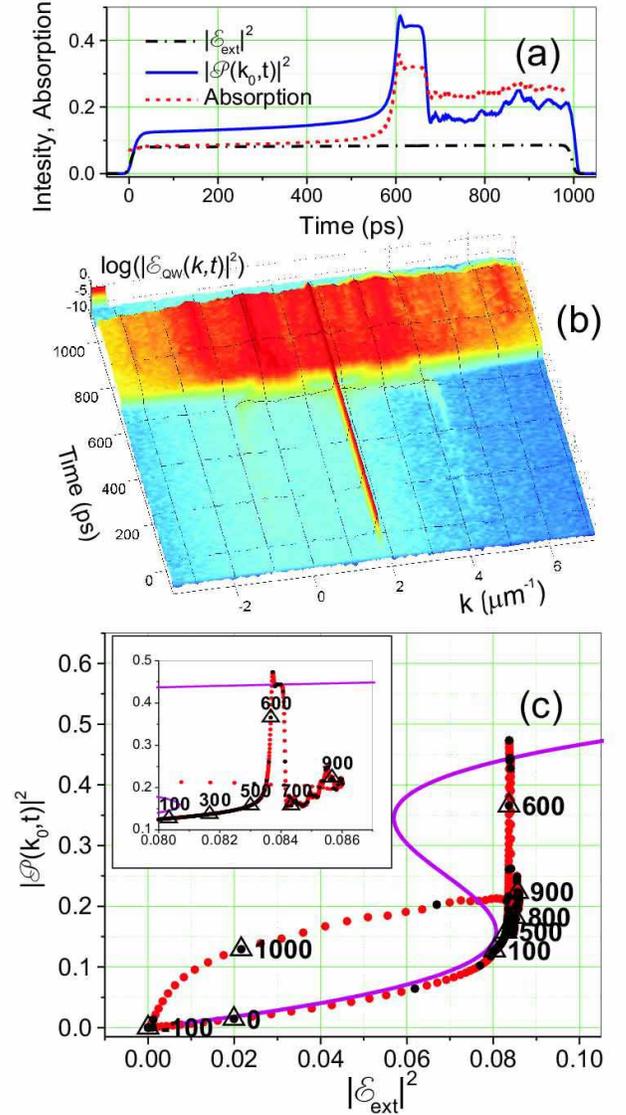}
\vspace{-1.cm} \caption{ \label{fig:4} The time dependences: of
input excitation pulse, calculated exciton polarisation and
absorption coefficient (\textit{a}); of angular spectra of the
scattered light (\textit{b}). Exciton polarisation vs pump
intensity during the excitation pulse (\textit{c}). Insert shows
the magnified part of the region of transitions. The time delay in
panel \textit{c} between red, black, and triangled dots is  1, 10,
and 100~ps, respectively; the triangles are labelled by time in
ps. Solid S-shaped curve is the solution of Eq.~(\ref{P0}).}
\vspace{-0.5cm}
\end{figure}

The numerical solution of Eqs.~(\ref{field},\ref{pol}) for long
($\approx 1$~ns) nearly rectangular excitation pulses demonstrates
a sharp threshold-like transition from a picture agreeing with the
described above stationary model to a completely different one,
compare Figs.~\ref{fig:3}\textit{a} and \textit{b}. Note that the
difference in the pump intensities between panels \textit{a} and
\textit{b} is $\approx 1$~\% only. The averaged intensity of
scattered noise increases above transition by several orders of
magnitude (note the logarithmic  vertical scale in
Fig.~\ref{fig:3}).

Figure~\ref{fig:4} explains the time kinetics of the scattering
above the threshold pump intensity. The shape of the excitation
pulse $\mathcal{E}_\mathrm{ext}^2$ (dash-dotted line in
Fig.~\ref{fig:4}\textit{a}) was chosen in such a way that the
threshold intensity is approached slowly during the pulse duration.
Actually two sharp transitions take place (at $t\approx 600$ and
700~ps). Both transitions are characterized by the jumps of the
driven mode amplitude $\mathcal{P}_0$ itself, see solid line in
Fig.~\ref{fig:4}\textit{a}.

In order to understand these transitions, we have to investigate
the stability of solutions of Eqs.~(\ref{field},\ref{pol}) in the
case of stationary external field $\mathcal{E}_\mathrm{ext}(t) =
\mathrm{const}$ with only one macroscopically filled mode, i.e.,
of the form $\mathcal{P}(k,t) = \tilde{\mathcal{P}}(k,t)
 + \delta_{k,k_0}\mathcal{P}_0 e^{- i \Omega_0 t}$,
and
$\mathcal{E}_\mathrm{QW}(k,t) = \tilde{\mathcal{E}}(k,t)
 + \delta_{k,k_0}\mathcal{E}_0 e^{- i \Omega_0 t}$.
Here
$|\tilde{\mathcal{P}}/\mathcal{P}_0|,~|\tilde{\mathcal{E}}/\mathcal{E}_0| \ll 1 $.
In zero order we get the following cubic equation for the amplitude of the
driven mode $\mathcal{P}_0$, \begin{equation}\label{P0}
\left[\Delta_0(\Delta_0+\Delta_{XC}) - A\beta \right] \mathcal{P}_0 -\Delta_0
F |\mathcal{P}_0|^2\mathcal{P}_0 = A\alpha \mathcal{E}_\mathrm{ext},
\end{equation} where $\Delta_0 = \hbar\Omega_0-E_\mathrm{MC}(k_0), \Delta_{XC}
= E_\mathrm{MC}(k_0) - E_X$.  Solution of
Eq.~(\ref{P0}) is shown as the solid S-shaped line in
Fig.~\ref{fig:4}\textit{c}. The S-shape is well known in the theory of
nonlinear cavities (see, e.g., in Refs.~\cite{Firth96,Kuszelewicz00}),
it brings a \textit{bistability} into the behaviour of the MC polaritons.
The influence of the bistability  was not analyzed in the existent theories of the
MC polariton scattering~\cite{Ciuti00,Ciuti01,Whittaker01,Savvidis01,Savasta03}.

As to the stability conditions, linearising Eqs.~(\ref{field},\ref{pol}),
we have to solve the linear eigenproblem for parametrically coupled signal
$\tilde{\mathcal{E}},\tilde{\mathcal{P}}(k,t) =
\tilde{\mathcal{E}},\tilde{\mathcal{P}}(k)e^{-i\omega t}$ and idler
$\tilde{\mathcal{E}}^*,\tilde{\mathcal{P}}^*(\bar{k},t)e^{-2i\Omega_0t} =
\bar{\tilde{\mathcal{E}}},\bar{\tilde{\mathcal{P}}}(\bar{k}) e^{-i\omega t}$
(where  $\bar{k} = 2k_0 - k$),
\begin{widetext}
\begin{eqnarray}\label{Pk}
\hbar\omega
 \begin{pmatrix}
    \tilde{\mathcal{E}}(k) \\
    \tilde{\mathcal{P}}(k) \\
   \bar{ \tilde{\mathcal{E}}}(\bar{k}) \\
   \bar{ \tilde{\mathcal{P}}}^(\bar{k})
  \end{pmatrix}
  =
  \begin{pmatrix}
   E_\mathrm{MC}(k)         &  \beta                         & 0                                   & 0\\
   A                        & E_X + 2 F |\mathcal{P}_0|^2    & 0                                   & F \mathcal{P}_0^2\\
   0                        & 0                              &2\Omega_0-E_\mathrm{MC}^*(\bar{k})   &-\beta^*  \\
   0                        & -(F\mathcal{P}_0^2)^*          &-A^*                                 & 2\Omega_0- E_X^* - 2 F^* |\mathcal{P}_0|^2
  \end{pmatrix}
  \begin{pmatrix}
    \tilde{\mathcal{E}}(k) \\
    \tilde{\mathcal{P}}(k) \\
   \bar{ \tilde{\mathcal{E}}}(\bar{k}) \\
   \bar{ \tilde{\mathcal{P}}}^(\bar{k})
  \end{pmatrix}
,
\end{eqnarray}
\end{widetext}
and check the sign of the imaginary parts of the eigenenergies.
Equation~(\ref{Pk}) appears to be  analogous to that
discussed in detail in Refs.~\cite{Ciuti01,Whittaker01,Savvidis01}.
We use the MC-X basis instead of
the UP-LP one. There is an advantage of the MC-X basis, because
the dependence of the interaction constant $F$  on $k$ can be
neglected. Then, in the space representation, it corresponds to
a local contact interaction $F \int |\mathcal{P}(x)|^4 dx$,
which allows to write a very efficient numerical code for
Eqs.~(\ref{field},\ref{pol}).

The idler and signal branches discussed above [see in
Figs.~\ref{fig:2}\textit{b},\ref{fig:3}] are the eigenvalues of
Eq.~(\ref{Pk}) calculated with $\mathcal{P}_0^2 = 0.14$, a threshold
value for the stimulated scattering into $k_s < 0, \,\,k_i > 2 k_0$.
And indeed, Fig.~\ref{fig:4}\textit{b} shows that at  $t < 600$~ps the
scattered noise is maximum at the predicted angles $k_s < 0, \,\,k_i >
2 k_0$. However, at $t \approx 600$~ps, instead of the stimulated
scattering into these modes, another instability develops. The
trajectory of the system on the $\left[|\mathcal{P}(k_0,t)|^2,
|\mathcal{E}_\mathrm{ext}(t)|^2\right]$ plane, see in
Fig.~\ref{fig:4}\textit{c}, shows that the instability at $t \approx
600$~ps is the jump of the $k_0$ mode between the lower and upper
branches of the S-shaped curve.

In the empty cavity with quadratic dispersion, the upper S' branch is
usually stable~\cite{Firth96,Kuszelewicz00}. In a MC with inflective
LP dispersion, it can become unstable against parametric scattering.
It can be seen from, e.g., Fig.~\ref{fig:2}\textit{c}, showing the
signal and idler branches calculated on the upper part of S-curve at
$|\mathcal{P}_0|^2 = 0.44$. Instead of damping, large gain is
realized for polariton modes just with  $k_s \gtrapprox 0, \,\,k_i
\lessapprox 2k_0$. As a result, a stimulated scattering into these
modes develops (see in Fig.~\ref{fig:4}\textit{b}). It is feed up by
the increased absorption, see dashed line in
Fig.~\ref{fig:4}\textit{a}. But the pumped mode becomes unstable and
eventually jumps   back into the lower position (at  $t \approx
700$~ps). Now, because we already have a well developed scattered
noise in the system, this lower position is more or less stable,
although subject to noisy fluctuations because of the already
developed scattered states and the competition between them and the
pumped mode.

We would like to note that within our theoretical model the
instability of the upper S' branch takes place in a narrower
region of parameters than in the experiment.  E.g., for the pump
at $\vartheta_0 = 14^\circ$ and $12.5^\circ$ (as in the experiment
in Fig.~\ref{fig:1}\textit{b}) it occurs for detuning
$0.1<\Delta<0.6$~meV and $0.15<\Delta<0.3$~meV, respectively.
However, the other scattering mechanisms, e.g., with phonons and
free carriers, which are neglected in our approach, can increase
substantially the region of the instability.

To conclude, we have found experimentally that above the MC
parametric scattering threshold the signal and idler maxima do not
shift along the renormalised LP branch but jump to  $k_s
\gtrapprox 0, \,\,k_i \lessapprox 2k_0$ directions. We show that
qualitatively this behavior can be explained via an interplay
between the the bistability of the driven mode itself and the
stimulated parametric scattering. The bistability, although well
known in empty nonlinear cavities, was not accounted for in the
existent theoretical models of the MC polaritons parametric scattering.
It acquires a new feature in MC with polariton resonance:
instability of the upper branch of the S-shaped dependence of
pumped mode polarisation vs pump. Although the exciton-exciton
scattering may be not the only mechanism causing the instability,
this feature is important and has to be taken into account.

\begin{acknowledgments}
We thank L.~V.~Keldysh for stimulating discussions,
and M.~S.~Skolnick  for samples and
discussions. This work was financially supported in part by the
Russian Foundation for Basic Research, Russian Ministry of Science,
and INTAS.
\end{acknowledgments}


\begin{thebibliography}{17}
\expandafter\ifx\csname natexlab\endcsname\relax\def\natexlab#1{#1}\fi
\expandafter\ifx\csname bibnamefont\endcsname\relax
  \def\bibnamefont#1{#1}\fi
\expandafter\ifx\csname bibfnamefont\endcsname\relax
  \def\bibfnamefont#1{#1}\fi
\expandafter\ifx\csname citenamefont\endcsname\relax
  \def\citenamefont#1{#1}\fi
\expandafter\ifx\csname url\endcsname\relax
  \def\url#1{\texttt{#1}}\fi
\expandafter\ifx\csname urlprefix\endcsname\relax\def\urlprefix{URL }\fi
\providecommand{\bibinfo}[2]{#2}
\providecommand{\eprint}[2][]{\url{#2}}

\bibitem[{\citenamefont{Savvidis et~al.}(2000)\citenamefont{Savvidis, Baumberg,
  Stevenson, Skolnick, Whittaker, and Roberts}}]{Savvidis00}
\bibinfo{author}{\bibfnamefont{P.~G.} \bibnamefont{Savvidis et~al.}},
  \bibinfo{journal}{Phys. Rev. Lett.}
  \textbf{\bibinfo{volume}{84}}, \bibinfo{pages}{1547} (\bibinfo{year}{2000}).


\bibitem[{\citenamefont{Houdr\'{e} et~al.}(2000)\citenamefont{Houdr\'{e},
  Weisbuch, Stanley, Oesterle, and Ilegems}}]{Houdre00}
\bibinfo{author}{\bibfnamefont{R.}~\bibnamefont{Houdr\'{e} et~al.}},
  \bibinfo{journal}{Phys. Rev. Lett.} \textbf{\bibinfo{volume}{85}},
  \bibinfo{pages}{2793–} (\bibinfo{year}{2000}).


\bibitem[{\citenamefont{Ciuti et~al.}(2000)\citenamefont{Ciuti, Schwendimann,
  Deveaud, and Quattropani}}]{Ciuti00}
\bibinfo{author}{\bibfnamefont{C.}~\bibnamefont{Ciuti et~al.}},
  \bibinfo{journal}{Phys. Rev. B} \textbf{\bibinfo{volume}{62}},
  \bibinfo{pages}{R4825} (\bibinfo{year}{2000}).


\bibitem[{\citenamefont{Tartakovskii et~al.}(2000)\citenamefont{Tartakovskii,
  Krizhanovskii, and Kulakovskii}}]{Tartakovskii00}
\bibinfo{author}{\bibfnamefont{A.~I.} \bibnamefont{Tartakovskii et~al.}},
  \bibinfo{journal}{Phys. Rev. B}
  \textbf{\bibinfo{volume}{62}}, \bibinfo{pages}{R13298}
  (\bibinfo{year}{2000}).


\bibitem[{\citenamefont{Stevenson et~al.}(2000)\citenamefont{Stevenson,
  Skolnick, Whittaker, Emam-Ismail, Tartakovskii, Savvidis, Baumberg, and
  Roberts}}]{Stevenson00}
\bibinfo{author}{\bibfnamefont{R.~M.} \bibnamefont{Stevenson et~al.}},
  \bibinfo{journal}{Phys. Rev. Lett.}
  \textbf{\bibinfo{volume}{85}}, \bibinfo{pages}{3680} (\bibinfo{year}{2000}).



\bibitem[{\citenamefont{Baumberg et~al.}(2000)\citenamefont{Baumberg, Savvidis,
  Stevenson, Tartakovskii, Skolnick, Whittaker, and Roberts}}]{Baumberg00}
\bibinfo{author}{\bibfnamefont{J.~J.} \bibnamefont{Baumberg et~al.}},
  \bibinfo{journal}{Phys. Rev. B}
  \textbf{\bibinfo{volume}{62}}, \bibinfo{pages}{R16247}
  (\bibinfo{year}{2000}).


\bibitem[{\citenamefont{Ciuti et~al.}(2001)\citenamefont{Ciuti, Schwendimann,
  and Quattropani}}]{Ciuti01}
\bibinfo{author}{\bibfnamefont{C.}~\bibnamefont{Ciuti et~al.}},
  \bibinfo{journal}{Phys. Rev. B} \textbf{\bibinfo{volume}{63}},
  \bibinfo{pages}{041303} (\bibinfo{year}{2001}).


\bibitem[{\citenamefont{Whittaker}(2001)}]{Whittaker01}
\bibinfo{author}{\bibfnamefont{D.~M.} \bibnamefont{Whittaker}},
  \bibinfo{journal}{Phys. Rev. B} \textbf{\bibinfo{volume}{63}},
  \bibinfo{pages}{193305} (\bibinfo{year}{2001}).


\bibitem[{\citenamefont{Savvidis et~al.}(2001)\citenamefont{Savvidis, Ciuty,
  Baumberg, Whittaker, Skolnick, and Roberts}}]{Savvidis01}
\bibinfo{author}{\bibfnamefont{P.~G.} \bibnamefont{Savvidis et~al.}},
  \bibinfo{journal}{Phys. Rev. B}
  \textbf{\bibinfo{volume}{64}}, \bibinfo{pages}{075311}
  (\bibinfo{year}{2001}).

\bibitem[{\citenamefont{Saba et~al.}(2001)\citenamefont{Saba, Ciuti, Bloch,
  Thierry-Mieg, Andre, Dang³, Kundermann, Mura, Bongiovanni, Staehli
  et~al.}}]{Saba01}
\bibinfo{author}{\bibfnamefont{M.}~\bibnamefont{Saba et~al.}},
 \bibinfo{journal}{Nature}
  \textbf{\bibinfo{volume}{414}}, \bibinfo{pages}{731} (\bibinfo{year}{2001}).


\bibitem[{\citenamefont{Savasta et~al.}(2003)\citenamefont{Savasta, Stefano,
  and Girlanda}}]{Savasta03}
\bibinfo{author}{\bibfnamefont{S.}~\bibnamefont{Savasta  et~al.}},
  \bibinfo{journal}{Phys. Rev. Lett.} \textbf{\bibinfo{volume}{90}},
  \bibinfo{pages}{096403} (\bibinfo{year}{2003}).

\bibitem[{\citenamefont{Huynh et~al.}(2003)\citenamefont{Huynh, Tignon,
  Larsson, Roussignol, Delalande, Andr\'{e}, Romestain, and Dang}}]{Huynh03}
\bibinfo{author}{\bibfnamefont{A.}~\bibnamefont{Huynh et~al.}},
  \bibinfo{journal}{Phys. Rev. Lett.} \textbf{\bibinfo{volume}{90}},
  \bibinfo{pages}{106401} (\bibinfo{year}{2003}).

\bibitem[{\citenamefont{Kulakovskii et~al.}(2001)\citenamefont{Kulakovskii,
  Tartakovskii, Krizhanovskii, Gippius, Skolnick, and
  Roberts}}]{Kulakovskii01-2}
\bibinfo{author}{\bibfnamefont{V.~D.} \bibnamefont{Kulakovskii et~al.}},
 \bibinfo{journal}{Nanotechnology}
  \textbf{\bibinfo{volume}{12}}, \bibinfo{pages}{475} (\bibinfo{year}{2001}).

\bibitem[{\citenamefont{Piermarocchi et~al.}(1996)\citenamefont{Piermarocchi,
  Tassone, Savona, Quattropani, and Schwendimann}}]{Piermarocchi96}
\bibinfo{author}{\bibfnamefont{C.}~\bibnamefont{Piermarocchi et~al.}},
  \bibinfo{journal}{Phys. Rev. B} \textbf{\bibinfo{volume}{53}},
  \bibinfo{pages}{15834} (\bibinfo{year}{1996}).

\bibitem[{\citenamefont{Malpuech et~al.}(1996)\citenamefont{Malpuech, Kavokin,
  Carlo, and Baumberg}}]{Malpuech02}
\bibinfo{author}{\bibfnamefont{G.}~\bibnamefont{Malpuech et~al.}},
  \bibinfo{journal}{Phys. Rev. B} \textbf{\bibinfo{volume}{65}},
  \bibinfo{pages}{153310} (\bibinfo{year}{1996}).

\bibitem[{\citenamefont{Firth and Scroggie}(1996)}]{Firth96}
\bibinfo{author}{\bibfnamefont{W.~J.} \bibnamefont{Firth}} \bibnamefont{and}
  \bibinfo{author}{\bibfnamefont{A.~J.} \bibnamefont{Scroggie}},
  \bibinfo{journal}{Phys. Rev. Lett.} \textbf{\bibinfo{volume}{76}},
  \bibinfo{pages}{1623–} (\bibinfo{year}{1996}).

\bibitem[{\citenamefont{Kuszelewicz et~al.}(2000)\citenamefont{Kuszelewicz,
  Ganne, Sagnes, Slekys, and Brambilla}}]{Kuszelewicz00}
\bibinfo{author}{\bibfnamefont{R.}~\bibnamefont{Kuszelewicz et~al.}},
  \bibinfo{journal}{Phys. Rev. Lett.} \textbf{\bibinfo{volume}{84}},
  \bibinfo{pages}{6006–} (\bibinfo{year}{2000}).

\end{thebibliography}

\end{document}